# Reconfigurable Hardware Implementation of the Successive Overrelaxation Method

Safaa J. Kasbah[1], Ramzi A. Haraty[1], and Issam W. Damaj[2]

## Introduction

*"Surely the first and oldest problems in every branch of mathematics spring from experience and are suggested by the world of external phenomena"* [1]. At the age of 2 or 3, we study addition by assembling collection of objects and counting them. At the age of 4 or 5, we start using an abstract mathematical construction, a model, known as the positive integers [2]. Later, in elementary school, we start using more complicated constructions known as operators. Apparently, these early taught mathematical constructions form the basis of our understanding of a variety of problems.

As mathematical problems become more complex, it might be still possible to find their solutions by means of available computing devices. However, there are several mathematical problems whose solutions are difficult to be realized using available computing power [3-5]. Examples of such problems are factoring very large numbers (*RSA* depends on this problem's computational difficulty) [5], finding the solution to partial differential equations [6], and deciding whether a knot in 3-dimensional Euclidean space is unknotted (topological problem) [7]. More of such problems can be found in [4-5, 8-10]. Complex problems in science and engineering, including the ones above, are computationally intensive in nature [11]. Factoring very large numbers can be achieved by computation only, since the underlying algorithmic procedures are well known. The same is true for solving partial differential equations or studying the topological unknotted problem and for thousands scientific and engineering endeavors. Several attempts to accelerate the computation of such complex mathematical problems have been motivated with the enormous advances in computing systems.

Many physical phenomena can be expressed as systems of linear equations. Numerical solutions for these equations allow us to glean valuable information about the system at hand. There are two basic approaches for solving linear systems: Direct Methods and Iterative Methods. In the first approach, a finite number of operations are performed to find the exact solution. In the second approach, an initial approximate of the solution is generated, then this initial guess is used to generate another approximate solution, which is more accurate than the previous one [12] The robustness of applying iterative methods over direct methods is shown in different areas including: circuit analysis and design, weather forecasting and analyzing financial market trends.

---

[1] Safaa J. Kasbah and Ramzi A. Haraty
Division of Computer Science and Mathematics, Lebanese American University, Beirut, Lebanon

[2] Issam W. Damaj
Department of Electrical and Computer Engineer, Dhofar University, Salalah, Sultanate of Oman

The well-known iterative methods are: *Gauss-Seidel*, *Multigrid*, *Jacobi* and Successive Over- Relaxation (*SOR)* which is of a interest in this chapter. *SOR* has been devised to accelerate the convergence of *Gauss-Seidel* and *Jacobi* [13], by introducing a new parameter,$\omega$, referred to as the relaxation factor. The *SOR* rate of convergence is highly dependent on the relaxation factor. The main difficulty of using *SOR* is finding a good estimate of the relaxation factor [12]. Several techniques have been proposed for determining the exact value of $\omega$ which accelerates the rate of convergence of the method [12, 13].

All available iterative methods packages, including *SOR*, are done in software. Examples are the: *ITPACK 3A*, *ITPACK 3B*, *ITPACK 2C*, *ITPACK 2D*, and the *ELLPACK* package [14, 15]. Several sequential and parallel techniques were used in these packages to accelerate the method [16].

The emergence of the new computing paradigm, Reconfigurable Computing (*RC*), introduces novel techniques for accelerating certain classes of applications including signal processing (e.g., weather forecasting, seismic data processing, Magnetic Resonance Imaging (*MRI*), adaptive filters), cryptography and *DNA* matching [17]. *RC*-systems combine the flexibility offered by software and the performance offered by hardware [18]. It requires a reconfigurable hardware, such as an *FPGA*, and a software design environment that aids in the creation of configurations for the reconfigurable hardware [17].

In [19], the first hardware implementation of an iterative method- the *Multigrid*- is presented. The speedup achieved demonstrates that hardware design can be suited for such computationally intensive applications. Toward proving the hypothesis that accelerated versions of the iterative methods can be realized in hardware, we undertook the first hardware implementation of the *SOR* method; using the same *FPGAs* that were used in [19-21].

In this chapter, we study the feasibility of implementing *SOR* in reconfigurable hardware. We use *Handel-C*, a higher-level design tool to code our design, which is analyzed, synthesized, and placed and routed using the *FPGAs* proprietary software (*DK* Design Suite, *Xilinx ISE 8.1i* and *Quartus II 5.1*). We target *Virtex II Pro*, *Altera Stratix* and *Spartan3L* which is embedded in the *RC10 FPGA*-based system from *Celoxica*. We report our timing results when targeting *Virtex II Pro* and compare them with a software version results written in *C++* and running on a General-Purpose Processor (*GPP*).

## Description of the Algorithm

The successive over-relaxation method is an iterative method used for finding the solution of elliptic differential equations. *SOR* has been devised to accelerate the convergence of *Gauss-Seidel* and *Jacobi* [14], by introducing a new parameter, $\omega$, referred to as the relaxation factor.

Given the linear system of equations:

$$A\varphi = b \tag{1}$$

the matrix $A$ can be written as

$$A = D + L + U \tag{2}$$

where $D$, $U$ and $L$ denote the diagonal, strictly upper triangular, and strictly lower triangular part of matrix $A$ [34].

Using the successive over relaxation technique, the solution of the *PDE* is obtained using:

$$x^{(k)} = (D - \omega L)^{-1}[\omega U + (1-\omega)D]x^{(k-1)} + \omega(D - \omega L)^{-1}b \tag{3}$$

where $x^{(k)}$ represents the $k^{th}$ iterate.

The *SOR* rate of convergence strongly depends on the choice of the relaxation factor, $\omega$ [3]. Extensive work has been done on finding a good estimate of this factor in the [0, 2] interval [3, 23].

Recent studies have shown that for the case where:

- $\omega = 1$: *SOR* simplifies to *Gauss-Seidel* method [24].

- $\omega \leq 1$ or $\omega \geq 2$ : *SOR* fails to converge [24].

- $\omega > 1$: *SOR* used to speedup convergence of a slow-converging process [34].

- $\omega < 1$: helps to establish convergence of diverging iterative process [23].

## Reconfigurable Computing

Today, it becomes possible to benefit from the advantages of both software and hardware with the presence of the *RC* paradigm [18] Actually, the first idea to fill the gap between the two computing approaches, hardware and software, goes back to the 1960s when Gerald Estrin proposed the concept of *RC* [23].

The basic idea of *RC* is the "ability to perform certain computations in hardware to increase the performance, while retaining much of the flexibility of a software solution" [18].

*RC*-systems can be either of fine-grained or of coarse-grained architecture. An *FPGA* is a fine-grained reconfigurable unit, while a reconfigurable array processor is a coarse-grained reconfigurable unit. In the fine-grained architecture each bit can be configured; while in the coarse-grained architecture, the operations and the interconnection of each processor can be configured. Example of a coarse-grained system is the *MorphoSys* which is intended for accelerating datapath applications by combining a *GPP* and an array of coarse-grained reconfigurable cells [2].

The realization of the *RC* paradigm is made possible by the presence of programmable hardware such as large scale Complex Programmable Logic Device (*CPLD*) and Field Programmable Gate Array (*FPGA*) chips [29]. *RC* involves the modification of the logic within the programmable device to suite the application at hand.

### Hardware compilation

There are certain procedures to be followed before implementing a design on an *FPGA*. First, the user should prepare his/her design by using either a schema editor or by using one of the Hardware Description Languages (*HDLs*) such as *VHDL* (Very high scale integrated circuit Hardware Description Language) and *Verilog*. With schema editors, the designer draws his/her design by choosing from the variety of available components (multiplexers, adders, resistors, ...) and connects them by drawing wires between them. Several companies supply schema editors where the designer can drag and drop symbols into a design, and clearly annotate each component [30]. Schematic design is considered simple and easy for relatively small designs. However, the emergence of big and complex designs has substantially decreased the popularity of schematic design while increasing the popularity of *HDL* design. Using an *HDL*, the designer has the choice of designing either the structure or the behavior of his/her design. Both *VHDL* and *Verilog* support structural and behavioral descriptions of the design at different levels of abstractions. In structural design, a detailed description of the system's components, sub-components and their interconnects are specified. The system will appear as a collection of gates and interconnects [26]. Though it has a great advantage of having an optimized design, structural presentation becomes hard, as the complexity of the system increases. In behavioral design, the system is considered as a black box with inputs and outputs only, without paying attention to its internal structure. In other words, the system is described in terms of how it behaves rather than in terms of its components and the interconnection between them. Though it requires more effort, structural representation is more advantageous than the behavioral representation in the sense that the designer can specify the information at the gate-level allowing optimal use of the chip area [27]. It is possible to have more than one structural representation for the same behavioral program.

Noting that modern chips are too complex to be designed using the schematic approach, we will choose the *HDL* instead of the schematic approach to describe our designs.

Whether the designer uses a schematic editor or an *HDL*, the design is fed to an Electronic Design Automation (*EDA*) tool to be translated to a netlist. The netlist can then be fitted on the *FPGA* using a process called place & route, usually completed by the *FPGA* vendors' tools. Then the user has to validate the place and route results by timing analysis, simulation and other verification methodologies. Once the validation process is complete, the binary file generated is used to (re)configure the *FPGA* device. More about this process is found in the coming sections.

Implementing a logic design on an *FPGA* is depicted in Fig. 1:

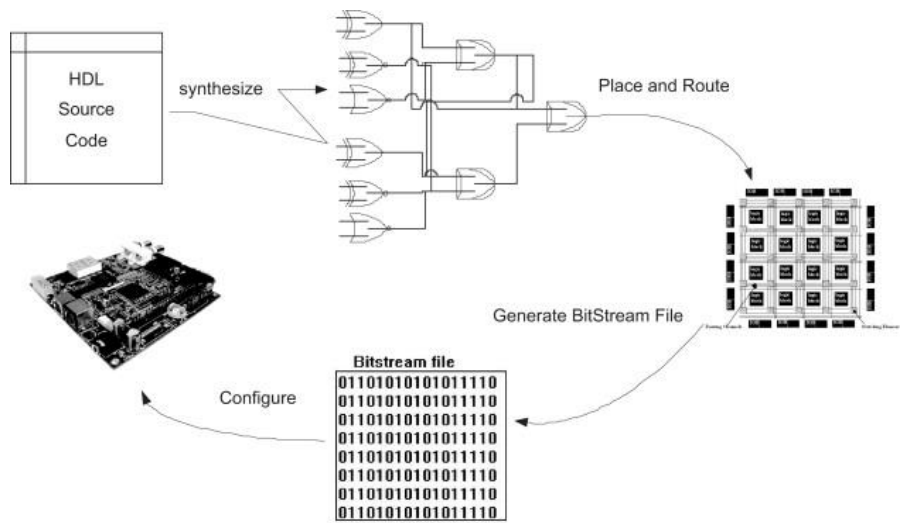

**Fig. 1.** *FPGA* design flow

The above process consumes a remarkable amount of time; this is due to the design that the user should provide using *HDL*, most probably *VHDL* or *Verilog*. The complexity of designing in *HDL*; which has been compared to the equivalent of assembly language; is overcome by raising the abstraction level of the design; this move is achieved by a number of companies such as *Celoxica*, *Cadence* and *Synopsys*. These companies are offering higher level languages with concurrency models to allow faster design cycles for *FPGA*s than using traditional *HDL*s. Examples of higher-level languages are *Handel-C*, *SystemC*, and *Superlog* [26].

### Handel-C Language

*Handel-C* is a high-level language for the implementation of algorithms on hardware. It compiles program written in a *C*-like syntax with additional constructs for exploiting parallelism [26]. The *Handel-C* compiler comes packaged with the *Celoxica DK Design Suite* which also includes functions and memory controller for accessing the external memory on the *FPGA*. A big advantage, compared to other *C* to *FPGA* tools, is that *Handel-C* targets hardware directly, and provides a few hardware optimizing features [8]. In contrast to other *HDLs*, such as *VHDL*, *Handel-C* does not support gate-level optimization. As a result, a *Handel-C* design uses more resources on an *FPGA* than a *VHDL* design and usually takes more time to execute. In the following subsections, we describe *Handel-C* features' that we have used in our design [28].

### Types and type operator

Almost all *ANSI-C* types are supported in *Handel-C* except for float and double. Yet, floating point arithmetic can still be performed using the floating-point library provided by *Celoxica*. Also, *Handel-C* supports all *ANSI-C* storage class specifier and type qualifiers except volatile and register which have no meaning in hardware. *Handel-C* offers additional types for creating hardware components such as memory, ports, buses and wires. *Handel-C* variables can only be initialized if they are global or if declared as static or const. *Handel-C* types are not limited to width since when targeting hardware, there is no need to be tied to a certain width. Variables can be of different widths, thus minimizing the hardware usage.

### par statement

The notion of time in *Handel-C* is fundamental. Each assignment happens in exactly one clock cycle, everything else is free [28]. An essential feature in *Handel-C* is the "par" construct which executes instructions in parallel.

### Handel-C targets

*Handel-C* supports two targets. The first is a simulator that allows development and testing of code without the need to use hardware, P1 in Fig 2. The second is the synthesis of a netlist for input to place and route tools which are provided by the *FPGA*'s vendors, P2 in Fig 2.

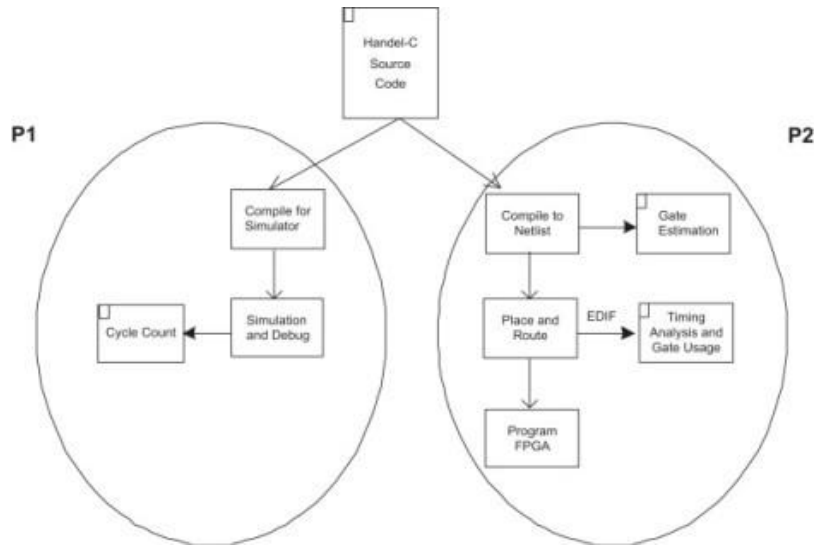

**Fig. 2**. Handel-C targets

The remaining of this section describes the phases involved in *P2*, as it is clear from *P1* that we can test and debug our design when compiled for simulation. The flow of the second target involves the following steps:

- **Compile to netlist**: The input to this phase is the source code. A synthesis engine, usually provided by the *FPGA* vendor, translates the original behavioral design into gates and flip flops. The resultant file is called the *netlist*. Generally, the *netlist* is in the Electronic Design Interchange Format (*EDIF*) format. An estimate of the logic utilization can be obtained from this phase.

- **Place and Route (PAR):** The input to this phase is the *EDIF* file generated from the previous phase; i.e. after synthesis. All the gates and flip flops in the netlist are physically placed and mapped to the *FPGA* resources. The *FPGA* vendor tool should be used to place and route the design. All design information regarding timing, chip area and resources utilization are generated and controlled for optimization at this phase.

- **Programming and configuring the FPGA:** After synthesis and place and route, a binary file will be ready to be downloaded into the *FPGA* chip [30, 31].

## Hardware Implementation of *SOR*

The successive over-relaxation method was designed using *Handel-C*, a higher-level hardware design tool. *Handel-C* comes packaged with *DK Design Suite* from *Celoxica*. It allows the designer to focus more on the specification of the algorithm rather than adopting a structural approach to coding [14]. *Handel-C* syntax is like the *ANSI-C* with additional extensions for expressing parallelism [14]. One of the most important features in *Handel-C* which is used in our implementation is the '*par*' construct that allows statements in a block to be executed in parallel and in the same clock cycle.

Our design has been tested using the *Handel-C* simulator; afterwards, we have targeted a *Xilinx Virtex II Pro FPGA*, an *Altera Stratix FPGA*, and *Spartan3L* which is embedded in an *RC10 FPGA*-based system from *Celoxica*. We have used the proprietary software provided by the devices vendors to synthesize, place and route, and analyze the design [28, 32, 33].

In Fig. 3 and Fig. 4, we present a parallel and a sequential version of *SOR*. In the first version, we used the '*par*' construct whenever it was possible to execute more than one instruction in parallel and in the same clock cycle without affecting the logic of the source code. The dots in the combined flowchart/concurrent process model which is shown in Fig. 3 represent replicated instances. Fig. 4 shows a traditional way of sequentially executing instructions on a general-purpose processor. Executing instructions in parallel have shown a substantial improvement in the execution of the algorithm.

To handle floating point arithmetic operations which are essential in finding the solution to *PDE* using iterative methods, we used the Pipelined Floating Point Library provided by *Celoxica* [28]. However, an unresolved bug in the current version of the *DK* simulator limited the usage of the floating-point operations to four in the design. The only possible way to avoid this failure was to convert/Unpack the floating-point numbers to integers and perform integer arithmetic on the obtained unpacked numbers. Though it costs more logic to be generated, the integer operations on the unpacked floating-point numbers have a minor effect on the total number of the design's clock cycles.

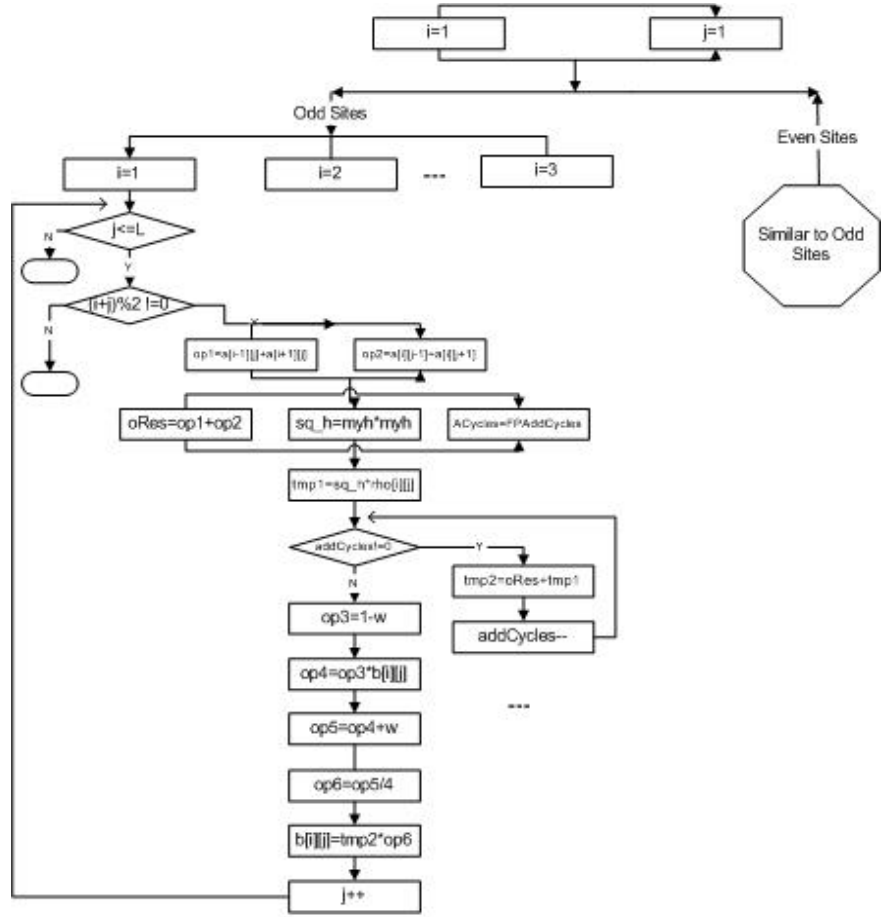

**Fig. 3.** *SOR* parallel version showing the combined flowchart concurrent process model. The dots represent replicated instances.

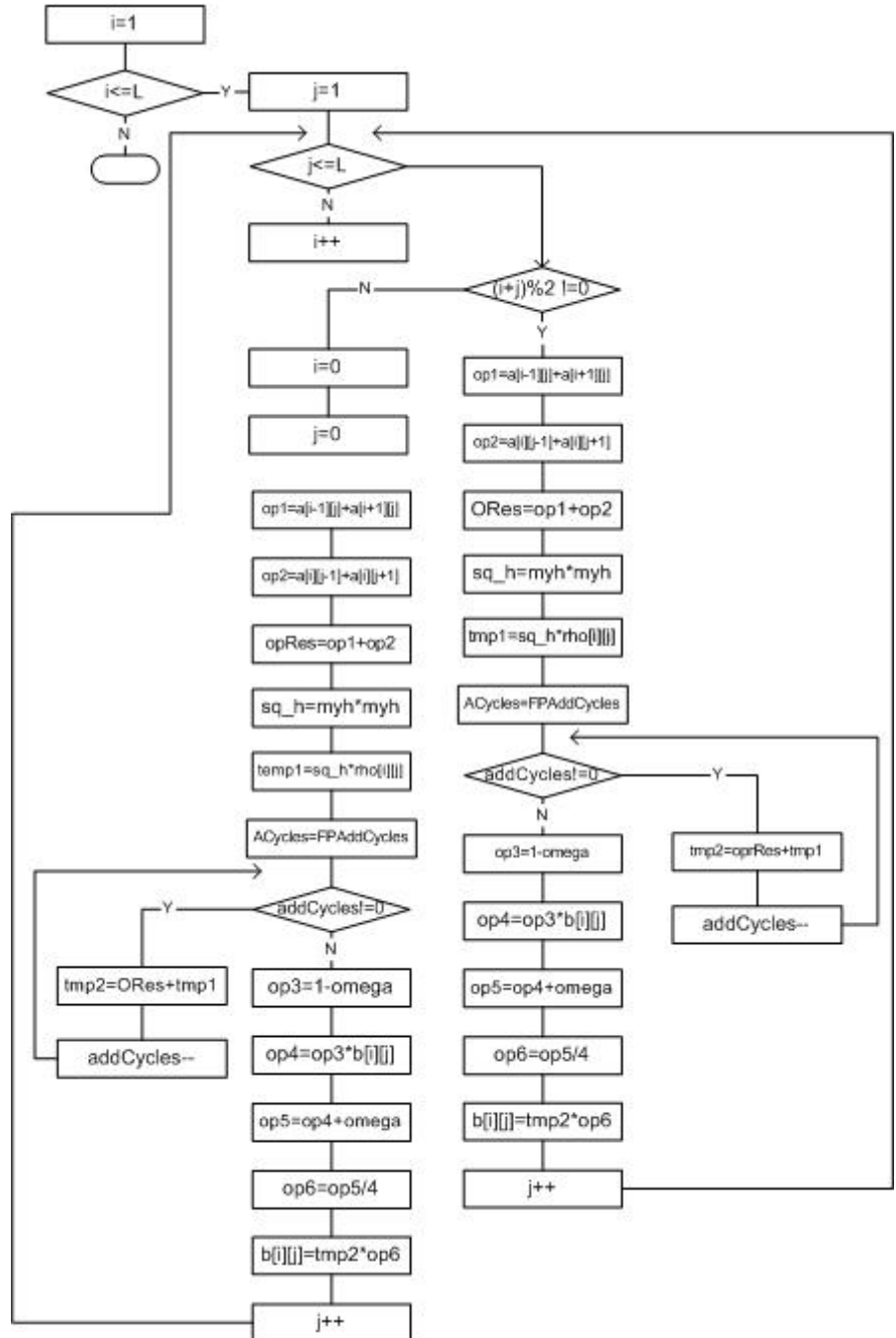

**Fig. 4.** *SOR* flowchart, sequential version

## Experimental Results

As mentioned before, the main objectives of this chapter are: i) studying the feasibility of implementing *SOR* method in hardware and ii) realizing an accelerated version of the method. The first objective is met by targeting high-performance *FPGAs*: *Virtex II Pro* (2vp7ff672-7), *Altera Stratix* (ep1s10f484c5), and *Spartan3L* (3s1500lfg320-4) which is embedded on *RC10* board from *Celoxica*. The second objective is met by comparing the timing results obtained, with a software version written in *C++* and compiled using *Microsoft Visual Studio .Net*. All the test cases were carried out on a Pentium (M) processor 2.0GHz, 1.99GB of *RAM*. The relaxation factor $\omega$ was chosen to be 1.5 [22]. The obtained results are based on the following criteria:

- **Speed of convergence:** the time it takes the SOR method to find the solution to the *PDE* in hand. In another word, it is the time needed to execute the *Multigrid* algorithm. In hardware implementation, the speed of convergence is measured using the clock cycles of the design divided by the frequency at which the design operates at. The first parameter is found using the simulator while the second is found using the timing analysis report which is generated using the *FPGA* vendor's tool.
- **Chip-area:** this performance criterion measures the number of occupied slices on the *FPGA* on which the design is implemented. The number of occupied slices is generated using the *FPGA* vendor's place and route tool.

We use the *FPGA* vendor's tools to analyze and report the performance results of each *FPGA*. The synthesis results obtained, for different problem sizes, when targeting *Virtex II Pro, Altera Stratix, and Spartan3L* are reported in Tables 1, 2 and 3, respectively.

**Table 1.** *Virtex II Pro* Synthesis Results

| Mesh Size | Occupied Slices | Total Equivalent Gate Count |
|---|---|---|
| 8x8 | 128 | 2918 |
| 16x16 | 136 | 3033 |
| 32x32 | 219 | 4807 |
| 64x64 | 265 | 5978 |
| 128x128 | 315 | 7125 |
| 256x256 | 610 | 14538 |
| 512x512 | 1098 | 23012 |
| 1024x1024 | 1601 | 31848 |
| 2048x2048 | 2289 | 53476 |

**Table 2.** *RC10 Spartan3L* Synthesis Results

| Mesh Size | Occupied Slices | Total Equivalent Gate Count |
|---|---|---|
| 8x8 | 302 | 279010 |
| 16x16 | 499 | 281001 |
| 32x32 | 589 | 282997 |
| 64x64 | 745 | 284000 |
| 128x128 | 877 | 285872 |
| 256x256 | 1201 | 297134 |
| 512x512 | 2010 | 299858 |

**Table 3.** *Altera Stratix* Synthesis Results

| Mesh Size | Total Logic Elements | Logic Element usage by nb. of LUT tables | Total Registers |
|---|---|---|---|
| 8x8 | 519 | 250 | 120 |
| 16x16 | 601 | 310 | 155 |
| 32x32 | 810 | 501 | 199 |
| 64x64 | 999 | 637 | 280 |
| 128x128 | 1274 | 720 | 347 |
| 256x256 | 1510 | 890 | 948 |
| 512x512 | 2286 | 1087 | 501 |
| 1024x1024 | 2901 | 1450 | 569 |
| 2048x2048 | 3286 | 1798 | 640 |

Fig 5 shows *SOR* execution time when targeting *Virtex II Pro FPGA* versus the execution time of *SOR* in *C++*. We started with a problem size of 8x8 and reached 2048x2048. Obviously, one can notice the acceleration of the method when moving from software implementation to hardware implementation. The speedup of the design, for different problem sizes, is shown in Table 4 and calculated as the ratio of Execution Time (*C++*) / Execution Time (*Handel-C*).

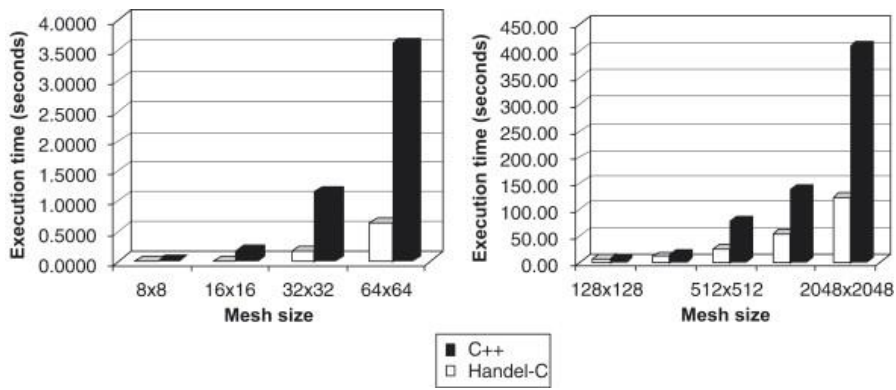

**Fig. 5** *SOR* execution time results in both versions, *C++* and *Handel-C*.

**Table 4.** The speedup of the design for different problem sizes

| Mesh Size | Speedup |
|---|---|
| 8x8 | 1.76 |
| 16x16 | 1.88 |
| 32x32 | 6.71 |
| 64x64 | 5.70 |
| 128x128 | 1.51 |
| 256x256 | 1.49 |
| 512x512 | 3.03 |
| 1024x1024 | 2.58 |
| 2048x2048 | 3.38 |

## Conclusion

In this chapter, we have studied the feasibility of implementing the *SOR* method on reconfigurable hardware. We used a hardware compiler, *Handel-C*, to code and implement our design which we map onto high-performance *FPGAs*: *Virtex II Pro*, *Altera Stratix*, and *Spartan3L* which is embedded in the *RC10* board from *Celoxica*. We used the *FPGAs* vendor's tool to analyze the performance of our hardware implementation. For testing purposes, we designed a software version of the algorithm and compiled it using *Microsoft Visual Studio .Net*. The software implementation results were compared to the hardware implementation results. The synthesis results prove that *SOR* is suitable for *FPGA* implementation; the timing results prove that *SOR* on hardware outperforms *SOR* on *GPP*. Soon, we plan to improve a) the speedup of the algorithm by designing a pipelined version of *SOR*; b) the efficiency of the algorithm by moving from *Handel-C* to a lower-level *HDL* such as *VHDL*. Besides, we will consider mapping the algorithm into a coarse grain reconfigurable system (e.g., MorphoSys) [34], and benefiting from the advantages of formal modeling [35]. We can extend the benefit of *SOR* by implementing other versions of the algorithm such as: Modified SOR (*MSOR*), Symmetric *SOR* (*SSOR*) and Unsymmetrical *SOR* (*USOR*).

## References

1. Hilbert, D. (1902). Mathematical problems. Bulletin of the American Mathematical Society, 8(10), 437-479.Gowers W, The Importance of Mathematics. Available at: http://www.dpmms.cam.ac.uk/~wtg10/importance.pdf
2. Bailey, D. H., & Borwein, J. M. (2005). Future Prospects for Computer-Assisted Mathematics. Notices of the Canadian Mathematical Society, 37(LBNL-59060).

14  Reconfigurable Hardware Implementation of the SOR Method3. Bailey, D., Borwein, P., & Plouffe, S. (1997). On the rapid computation of various polylogarithmic constants. Mathematics of Computation of the American Mathematical Society, 66(218), 903-913.
4. Bailey, D. H., Borwein, J. M., Kapoor, V., & Weisstein, E. W. (2006). Ten problems in experimental mathematics. The American Mathematical Monthly, 113(6), 481-509.
5. Morton, K. W., & Mayers, D. F. (2005). Numerical solution of partial differential equations: an introduction. Cambridge university press.
6. Burde, G and Zieschang, H (1985) Knots Walter de Gruyter Studies in Mathematics, Berlin
7. Hass, J., Lagarias, J. C., & Pippenger, N. (1999). The computational complexity of knot and link problems. Journal of the ACM (JACM), 46(2), 185-211.
8. Havas, G. (2003). On the complexity of the extended Euclidean algorithm. Electronic Notes in Theoretical Computer Science, 78, 1-4.
9. Havas G. and Seifert J P (1999) The complexity of the extended GCD problem. Mathematical foundations of computer science, Lecture Notes in Computer Science. pp 103-113
10. DeMillo, R. A., & Lipton, R. J. (1979, April). Some connections between mathematical logic and complexity theory. In Proceedings of the eleventh annual ACM symposium on Theory of computing (pp. 153-159). ACM.
11. Young, D. (1954). Iterative methods for solving partial difference equations of elliptic type. Transactions of the American Mathematical Society, 76(1), 92-111.
12. Evans G, Blackledge J and Yardley P (2000) Numerical Methods for Partial Differential Equations. Springer-Verlag, London
13. Bailey W (2003) The Successive Over Relaxation Algorithm and its application to Numerical Solutions of Elliptic Partial Differential Equations. B.Sc. project, Dublin Institute of Technology
14. Kincaid, D. R. (2004). Celebrating fifty years of David M. Young's successive overrelaxation method. In Numerical mathematics and advanced applications (pp. 549-558). Springer, Berlin, Heidelberg.
15. Zarka C, Edward G, and Freedman C (1990) Efficient Decomposition and Performance of Parallel PDE, FFT, Monte Carlo Simulations, Simplex, and Sparse Solvers. Proceedings of the 1990 ACM/IEEE conference on Super Computing. pp 455-464
16. Li, Y., Callahan, T., Darnell, E., Harr, R., Kurkure, U., & Stockwood, J. (2000, June). Hardware-software co-design of embedded reconfigurable architectures. In Proceedings of the 37th Annual Design Automation Conference (pp. 507-512). ACM.
17. Compton K and Hauck S (2002) Reconfigurable Computing: A Survey of Systems and Software. In: ACM Computing Surveys. 34: 2: pp. 171-210
18. Kasbah, S. J., Haraty, R. A., & Damaj, I. W. (2008). Reconfigurable Hardware Implementation of the Successive Overrelaxation Method. In Advances in Industrial Engineering and Operations Research (pp. 453-466). Springer, Boston, MA.

19. Kasbah S and Damaj I (2006) A hardware implementation of Multigrid Algorithms. Poster Session: 17th International Conference on Domain Decomposition Methods, Austria
20. Kasbah, S. J., Damaj, I. W., & Haraty, R. A. (2008). Multigrid solvers in reconfigurable hardware. Journal of Computational and Applied Mathematics, 213(1), 79-94.
21. Kulsrud H E (1961) A practical technique for the determination of the optimum relaxation factor of the successive over-relaxation method. Communications of the ACM. 4: 4: pp.184-187
22. Vahid F and Givargis T (2002). Embedded systems design: a unified hardware/software introduction. Wiley, New York
23. Lee, M. H., Singh, H., Lu, G., Bagherzadeh, N., Kurdahi, F. J., Eliseu Filho, M. C., & Alves, V. C. (2000). Design and implementation of the MorphoSys reconfigurable computing processor. Journal of VLSI signal processing systems for signal, image and video technology, 24(2-3), 147-164.
24. Todman T J ,Constantinides G A , Wilton S J E, Mencer O Luk W and Cheung P.Y.K. (2005) Reconfigurable computing: architectures and design methods. IEE Proceedings: Computers and Digital Techniques. 152: 2: pp193-197
25. Turely J (2003) How Chips are Designed Prentice Hall, Professional Technical Reference
26. Valentina S K (2004) Designing A Digital System with VHDL. Academic Open Internet Journal. vol: 11
27. Celoxica (2007) www.celoxica.com
28. Peter C (2000). Overview: Hardware Compilation and the Handel-C language. Oxford University Computing Laboratory; http://web.comlab.ox.uk/oucl/work/christian .peter/overview_handelc.html
29. Cong J (1997) FPGAs Synthesis and Reconfigurable Computing. University of California, Los Angeles: http://www.ucop.edu/research/micro/96_97/96_176.pdf
30. Shewel J (1998) A Hardware / Software Co-Design System using Configurable Computing Technology.
http://ipdps.cc.gatech.edu/1998/it/schewel.pdf
31. Altera Inc. (2007) www.altera.com
32. Xilinx (2007) www.xilinx.com
33. Damaj, I., & Diab, H. (2003). Performance analysis of linear algebraic functions using reconfigurable computing. The Journal of Supercomputing, 24(1), 91-107.
34. Damaj, I., Hawkins, J., & Abdallah, A. (2003, July). Mapping high-level algorithms onto massively parallel reconfigurable hardware. In IEEE International Conference of Computer Systems and Applications (pp. 14-22).